\documentclass[letterpaper,12pt]{article} 
\usepackage{amsmath,amssymb}

\providecommand{\e}{\ensuremath{\mathrm{e}}}
\providecommand{\bvec}[1]{\ensuremath{\mathbf{#1}}}
\providecommand{\kint}{\int \frac{d^d k}{(2\pi)^d}\;}

\author{W.~M\"uck\thanks{Email address: \texttt{wmueck@sfu.ca}}\
~and K.~S.~Viswanathan\thanks{Email address: \texttt{kviswana@sfu.ca}}\\
\emph{\small Department of Physics, Simon Fraser University, Burnaby,
British Columbia, V5A1S6 Canada}}
\title{A Regularization Scheme for the AdS/CFT Correspondence}

\begin{document}
\maketitle
\begin{abstract}
The prescription of the AdS/CFT correspondence is refined by
using a regularization procedure, which makes it possible to
calculate the divergent local terms in the CFT two-point
function. We present the procedure for the example of the scalar
field.  
\end{abstract}

\newpage
It has been stated in most papers on this subject that the
correspondence between a fields theory on anti-de Sitter space (AdS)
and a conformal field theory (CFT) on its boundary is formally
described by the formula \cite{Gubser,Witten} 
\begin{equation}
\label{form}
  \e^{-I_{AdS}[\phi]} = \left\langle \exp \int d^dx\; \phi_0(x)
  \mathcal{O}(x) \right\rangle, 
\end{equation}
where the action $I_{AdS}$ is calculated on-shell for a field
configuration $\phi$ satisfying a Dirichlet condition on the AdS
boundary and the boundary value $\phi_0$ couples as a current to the
conformal field $\mathcal{O}$ living on the AdS boundary. Thus, the formula
\eqref{form} enables one to calculate correlation functions of the
field $\mathcal{O}$ in the boundary conformal field theory. 
There are two subtle points, which this formal
description does not address. First, a Dirichlet boundary value
problem in general is not well defined for anti-de Sitter space, since
a generic field does not propagate to the boundary. This point has been
addressed by formulating the theory with a boundary lying inside the
anti-de Sitter space. Using this approach the two-, three- and
four-point functions of various fields have been calculated
(see \cite{Freedman4} for a recent comprehensive list of references). 
Secondly, although the $\epsilon$-prescription yields the 
non-contact contributions to the correlators in agreement with
conformal field theory, the singular contributions for coincidence points
have so far escaped a direct calculation. Previous work has focused on
obtaining a finite action by interpreting the correlators as
distributions \cite{Arefeva} and regularizing the action by adding
boundary counter terms \cite{Chalmers,Emparan}. Two loop corrections
to Super Yang Mills correlators have been calculated in
\cite{Gonzalez} and contribute only to the contact terms. 
However, we think that a regularization scheme
must include a prescription on how to calculate the divergent
contact contributions before any counterterms are added. 

The aim of this letter is to provide a refined prescription of the
AdS/CFT correspondence, with which non-local and local terms of the CFT
two-point function can be calculated. We would like to emphasize that it
is not our aim to obtain regularized CFT correlation functions. 

We shall use the conventional representation of anti-de Sitter space
by the space $x^i \in \mathbb{R}$, ($i=1,\ldots d$), $x^0>0$ with the
metric
\begin{equation}
\label{metric}
  ds^2 = (x^0)^{-2} dx^\mu dx^\mu.
\end{equation}
Let us consider the example of a scalar field \cite{Mueck1}, 
which satisfies the equation of motion
\begin{equation}
\label{eqmot}
  (\nabla^2-m^2)\phi(x) = \left[ x_0^2 \partial_\mu \partial_\mu -
  x_0(d-1) \partial_0 - m^2 \right] \phi(x) =0. 
\end{equation}
A solution of equation \eqref{eqmot} can be written in the form
  \[ \phi(x) = \int d^d y K(x,\bvec{y}) \phi_\epsilon(\bvec{y}), \]
where $\phi_\epsilon(\bvec{y})$ is some boundary field and the bulk-boundary
kernel is given by
\begin{equation}
\label{regkernel}
  K(x,\bvec{y}) = \kint \left(\frac{x_0}{\epsilon}
  \right)^\frac{d}2
  \frac{K_\alpha(kx_0)}{K_\alpha(k\epsilon)}
  \e^{-i\bvec{k} \cdot(\bvec{x}-\bvec{y}) -\mu k^2}.
\end{equation}
Here, $K_\alpha$ is a modified Bessel function (Mac Donald function)
and $\alpha$ is related to the mass parameter by
  \[ \alpha= \sqrt{\frac{d^2}{4}+m^2}. \]
In addition, we have introduced the regulating factor $\e^{-\mu k^2}$
in order to make the integral well defined for all values of $x_0$ and
$\bvec{x}-\bvec{y}$. In the limit $\mu\to0$ equation \eqref{regkernel}
reduces to the standard Dirichlet kernel. It will turn out that the
limits $\mu\to0$ and $\epsilon\to0$ should be taken simultaneously
with $\mu$ being of order $\epsilon^2$. 

The CFT two-point function is determined by the boundary normal
derivative of the kernel \eqref{regkernel}, which is given by
\begin{equation}
\label{delK}
  \partial_0 K(x,\bvec{y})|_\epsilon = \frac1{\epsilon} \kint
  \left\{ \frac{d}2-\alpha + k\frac{\partial}{\partial k} \ln \left[
  (k\epsilon)^\alpha K_\alpha (k\epsilon)\right]\right\}
  \e^{-i\bvec{k}\cdot(\bvec{x}-\bvec{y})-\mu k^2}.
\end{equation}
Consider first the case of non-coincident points $\bvec{x}$ and
$\bvec{y}$. As without regularization, we use the expansion 
\begin{equation}
\label{kexpand} 
  z^\alpha K_\alpha(z) = 2^{\alpha-1} \Gamma(\alpha) \left[ 1 +
  \sum_{j=1}^\infty \frac1{j! (1-\alpha)_j} 
  \left(\frac{z}2\right)^{2j}
  - \frac{\Gamma(1-\alpha)}{\Gamma(1+\alpha)} 
  \left(\frac{z}2\right)^{2\alpha}
  \sum_{j=0}^\infty \frac1{j! (1+\alpha)_j} 
  \left(\frac{z}2\right)^{2j} \right],
\end{equation}
where we used the notation
$(a)_j=\Gamma(a+j)/\Gamma(a)$. Proceeding to expand the
logarithm in equation \eqref{delK} one obtains 
\begin{equation}
\label{expand}
  z \frac{\partial}{\partial z} \ln \left[
  z^\alpha K_\alpha(z) \right]
  = \frac{z^2}{2(1-\alpha)} +\cdots 
  - \frac{\Gamma(1-\alpha)}{\Gamma(1+\alpha)} 2^{1-2\alpha} \alpha
  z^{2\alpha} +\cdots,
\end{equation}
where the first set of dots indicates analytic terms of order $z^{2n}$
($n>1$) and the second set higher order non-analytic
terms. Substituting equation \eqref{expand} into equation \eqref{delK}
we recognize integrals of the type 
\begin{align}
\notag
  \kint k^\beta \e^{-i\bvec{k}\cdot\bvec{x}-\mu k^2} &= 
  \frac{|\bvec{x}|^{1-\frac{d}2}}{(2\pi)^{\frac{d}2}} 
  \int\limits_0^\infty
  dk\; k^{\frac{d}2+\beta} J_{\frac{d}2-1}(k |\bvec{x}|)\, \e^{-\mu k^2} \\
\label{integral}
  &= \frac{\Gamma\left(\frac{d+\beta}2\right)}{2^d\pi^\frac{d}2
  \Gamma\left(\frac{d}2\right) \mu^\frac{d+\beta}2}
  \Phi\left( \frac{d+\beta}2; \frac{d}2; -\frac{|\bvec{x}|^2}{4\mu}\right).
\end{align}
Here, $\Phi(a,c,z)$ is the degenerate hypergeometric function. For
$\bvec{x}\not= 0$ we can take the $\mu\to 0$ limit and replace $\Phi$
with the leading term of its asymptotic expansion \cite{Seaborn},
which yields   
\begin{equation}
\label{asymint}
  \kint k^\beta \e^{-i\bvec{k}\cdot\bvec{x}-\mu k^2}
  \overset{\mu\to0}{=}
  \frac{2^\beta\Gamma\left(\frac{d+\beta}2\right)}{\pi^\frac{d}2
  \Gamma\left(-\frac\beta2\right)} \frac1{|\bvec{x}|^{d+\beta}}.
\end{equation}
We notice that for the analytic terms, $\beta=2n$, the gamma function in
the denominator diverges, so the integral becomes zero. The same holds
for the subleading terms of the asymptotic expansion. Hence, the
analytic terms do not contribute to the finite distance two-point
function. On the other hand, for the leading non-analytic term,
$\beta=2\alpha$, we obtain (still for $\bvec{x}\not= \bvec{y}$)
\begin{equation}
\label{delKfinal}
  \partial_0 K(x,\bvec{y})|_\epsilon \overset{\mu\to0}{=} 2\alpha
  c_\alpha \frac{\epsilon^{2\alpha-1}}{|\bvec{x}-\bvec{y}|^{2\Delta}},
\end{equation}
where $\Delta=d/2+\alpha$ and $c_\alpha =
\Gamma(\Delta)/(\pi^{d/2} \Gamma(\alpha))$. The higher order
non-analytic terms in equation \eqref{expand} contribute with higher
powers of $\epsilon$ and can be neglected in the $\epsilon\to0$
limit. Hence, we find agreement with previous results
\cite{Mueck1}. A nice and simple check of this procedure is
provided by the conformally coupled scalar field, $\alpha= 1/2$,
where the calculation can be done exactly.    

Let us now turn to the local terms in the two-point
function. In this case, it is more useful not to carry out the expansion 
\eqref{expand}, since all its terms would contribute because of
$\Phi(a,c,0)=1$. Consider instead the expression \eqref{delK} directly
with $\bvec{x}=\bvec{y}$. Using the identity 
 \[  z\frac\partial{\partial z} \ln [z^\alpha K_\alpha(z)] = -z
  \frac{K_{\alpha-1}(z)}{K_\alpha(z)} \]
and changing the integration variable to $\bvec{s}= \sqrt{\mu}
\bvec{k}$ we find
\begin{equation}
\label{loc1}
  \partial_0 K(x,\bvec{x})|_\epsilon =
  \frac1{\epsilon(4\pi\mu)^\frac{d}2} \left[ \frac{d}2-\alpha 
  - \frac{2a}{\Gamma\left(\frac{d}2\right)} \int ds\; s^d \e^{-s^2} 
  \frac{K_{\alpha-1}(as)}{K_\alpha(as)} \right],
\end{equation}
where the parameter $a$ denotes the ratio
$a=\epsilon/\sqrt{\mu}$. The $s$-integral is well-defined for all
positive $a$, although in general not elementary. However, we know
that it represents some positive number. Hence, we obtain
\begin{equation}
\label{loc2} 
  \partial_0 K(x,\bvec{x})|_\epsilon = -
  \frac\gamma{\epsilon(4\pi\mu)^\frac{d}2},
\end{equation} 
where $\gamma$ is a regularization dependent parameter satisfying
  \[ \alpha-\frac{d}2 < \gamma < \infty.\]
Moreover, if
$2\alpha<d$, i.e.\ for $-d^2<4m^2<0$, one could determine $a$ such
that $\gamma=0$. For example, in the
conformally coupled case, $\alpha=1/2$, one finds 
  \[a_\frac12 = \frac{(d-1)\Gamma\left(\frac{d}2\right)}{
  2\Gamma\left(\frac{d+1}2\right)}. \] 

The contribution \eqref{loc2} to the two-point function can be 
compensated by adding a local counterterm 
\begin{equation}
\label{count}
  I_{c} = - \frac\gamma2 \int d^d x\; \epsilon^{-d}
  [\phi_\epsilon(\bvec{x})]^2
\end{equation}
to the AdS action of a scalar field. However, one must notice that
this only eliminates the contact term, but does not regularize the
two-point function.  

In conclusion, we have presented in this letter a refined prescription
of the AdS/CFT correspondence, which invokes a regularization scheme
in order to calculate the contact contributions to the CFT
correlators. Our calculation shows that in general there is such a
contribution to the two-point function, but it can be compensated by
adding a covariant local surface term to the AdS
action. Alternatively, it might serve as a regulator for the CFT
two-point function and thus eliminate the need to add counterterms. 
We think that this is an interesting topic for further study. 
   
This work was supported in
part by a grant from NSERC. W.\ M.\ is grateful to Simon Fraser
University for financial support.

\end{document}